\documentclass[twocolumn]{aastex631}

\def\OIII{[\mbox{O\,{\sc iii}}]$\lambda 5007$}

\def\NII{[\mbox{N\,{\sc ii}}]$\lambda 6584$}

\def\Ha{{H$\alpha$}}
\def\Hb{{H$\beta$}}

\def\NIIHa{[\mbox{N\,{\sc ii}}]$\lambda 6583$/H$\alpha$}

\def\OIIIHb{[\mbox{O\,{\sc iii}}]$\lambda 5007$/H$\beta$}

\def\LOIIIs4{$L[\mbox{O\,{\sc iii}}]$/$\sigma^4$}

\def\Msunyr{${\rm M_{\odot}yr^{-1}}$}

\def\ergs{${\rm erg}~{\rm s}^{-1}$}
\def\kms{${\rm km}~{\rm s}^{-1}$}
\newcommand{\ergcms}	{\ifmmode {\rm erg\,cm}^{-2}\,{\rm s}^{-1} \else erg\,cm$^{-2}$\,s$^{-1}$\fi}

\usepackage{graphicx}
\usepackage{sistyle}
\SIthousandsep{,}

\newcommand{\Nall}{\num{124116}}				
\newcommand{\Nemission}{\num{71157}}				
\newcommand{\Nweak}{\num{44055}}			
\newcommand{\Nnonemission}{\num{8904}}		
\newcommand{\NSF}{\num{59395}}				
\newcommand{\Ncomp}{\num{8099}}			
\newcommand{\NLINER}{\num{1652}}			
\newcommand{\NSY}{\num{2011}}				

\newcommand{\NtrainingSF}{\num{11879}}		

\newcommand{\NtestSFTP}{\num{13922}}		
		
\newcommand{\NtestSFhzTP}{\num{621}}		
\newcommand{\Ntestemi}{\num{14232}} 		
\newcommand{\Ntestalllow}{\num{45402}}			
\newcommand{\Ntestallhi}{\num{23026}}	

\newcommand{\NOSSY}{\num{664187}}

\begin{document}

\title{Capturing star formation activity from compressed photometric images of galaxies}

\correspondingauthor{Kyuseok Oh}
\email{oh@kasi.re.kr}

\author[0000-0002-5037-951X]{Kyuseok Oh}
\affiliation{Korea Astronomy and Space Science Institute, Daedeokdae-ro 776, Yuseong-gu, Daejeon 34055, Republic of Korea}

\author{M. Dennis Turp}
\affiliation{Snowfinch AI GmbH, Birchstrasse 185, 8050 Z{\"u}rich, Switzerland}



\begin{abstract}
We present a novel approach for classifying star-forming galaxies using photometric images. 
By utilizing approximately $124,000$ optical color composite images and spectroscopic data of 
nearby galaxies at $0.01<z<0.06$ from the Sloan Digital Sky Survey, 
along with follow-up spectroscopic line measurements from the OSSY catalog, 
and leveraging the Vision Transformer machine-learning technique, 
we demonstrate that galaxy images in JPEG format alone can be directly used to determine 
whether star-forming activity dominates the galaxy, 
bypassing traditional spectroscopic analyses such as emission-line diagnostic diagrams. 
We anticipate that this method holds significant potential for application in current and future large-scale surveys, 
such as Euclid, the Dark Energy Survey (DES), and the Legacy Survey of Space and Time (LSST).
\end{abstract}



\section{Introduction} \label{sec:intro}
The physical state of galaxies and their classification have been among the most crucial 
and fundamental topics in astrophysics. A Variety of spectroscopic properties have been used 
to unveil which physical mechanisms govern host galaxies.

The shape of the continuum in a galaxy’s spectral energy distribution (SED) broadly reveals 
the types of stellar population that compose the galaxy \citep{Strateva01, Baldry04, Blanton09}. 
Early-type galaxies exhibit a rising continuum toward longer wavelengths in the optical band, 
with little or no ultraviolet (UV) light and a strong 4000 \AA\ break. 
In contrast, a bluer continuum with prominent UV emission and a weak 4000 \AA\ break is 
characteristic of late-type galaxies \citep{Balogh99, Kauffmann03intro, Kaviraj07}.

The presence of absorption lines is also known to distinguish different galaxy populations: 
Ca H\&K and Mg b lines dominate the spectra of old and passive early-type galaxies, 
while Balmer absorption lines (e.g., H$\delta$ and H$\gamma$) are indicative of star-forming galaxies 
\citep{Worthey94, Trager00}.

Along with these methods, emission-line ratios and diagnostic diagrams proposed by \citet[][hereafter referred to as BPT diagrams]{Baldwin81} enable us to investigate galaxy spectral classifications not only for individual objects but also for large samples of galaxies using spectroscopic surveys such as the Sloand Digital Sky Survey (SDSS, \citealt{York00}).

BPT diagrams have been extensively used to diagnose the physical state of ionized gas 
in active galactic nuclei (AGN) that host supermassive black holes (SMBHs) at their centers. 
Unlike star-forming galaxies, AGN produces high-energy photons in the ultraviolet (UV) and X-ray regimes 
as matter is accreted onto the SMBH. 
Highly penetrating UV photons and X-rays can reach deeper into the gas clouds 
surrounding the narrow-line region 
(NLR, typically spanning hundreds to a thousand parsecs in Seyfert galaxies), 
partially ionizing hydrogen and generating extended partially ionized regions 
with significant collisional excitation \citep{Osterbrock06}. 
This process naturally leads to elevated ratios of collisionally-excited forbidden lines 
relative to the Balmer emission lines produced by photoionization, 
thereby enabling the distinction between star-forming galaxies and AGNs.

Since \citet{Baldwin81} first introduced these diagnostics, 
the classification schemes have been progressively refined to account for 
composite objects (exhibiting both star-forming and AGN characteristics), 
Seyferts, and low-ionization nuclear emission-line regions (LINERs) 
\citep{Heckman80, Veilleux87, Kewley01, Kauffmann03, Kewley06, Schawinski07}. 
For this reason, obtaining high-quality spectroscopic data and 
accurately measuring emission-line strengths and ratios 
have become central efforts in many areas of modern astrophysics.

Owing to the revolutionary work achieved by SDSS a couple of decades ago, 
spectroscopic data for millions of galaxies, along with photometric data, 
are publicly available (e.g., \citealt{Abazajian09}). 
However, spectroscopic observations and data acquisitions remain expensive and 
time-consuming compared to acquiring photometric data. 
Despite their importance in studying the essential properties of galaxies, 
the uneconomic nature of spectroscopic observations has been a major issue, 
especially for large samples of galaxies. 
In the era of the next-generation ground-based surveys, 
such as the Legacy Survey of Space and Time (LSST, \citealt{Ivezic19}), 
which is expected to break new ground in our understanding of galaxy formation and evolution, 
this challenge persists.   

Recent advances in machine-learning techniques and their implications for astrophysics, along with increased computing power, 
have marked a turning point in addressing astrophysical questions: 
recovering detailed features of galaxies in photometric images \citep{Schawinski17},
decomposing AGN components from host galaxy photometric images \citep{Stark18}, 
examining the quenching and evolution of star formation in galaxies \citep{Schawinski18}, 
estimation of photometric redshifts \citep{Lee21}, 
and classifying galaxy morphology \citep{Dieleman15, DominguezSanchez18, Cheng20, Walmsley20, Gupta22, Kang22, Tarsitano22, Seo23}.

\begin{figure*}
\centering
\includegraphics[width=0.82\linewidth, angle=0]{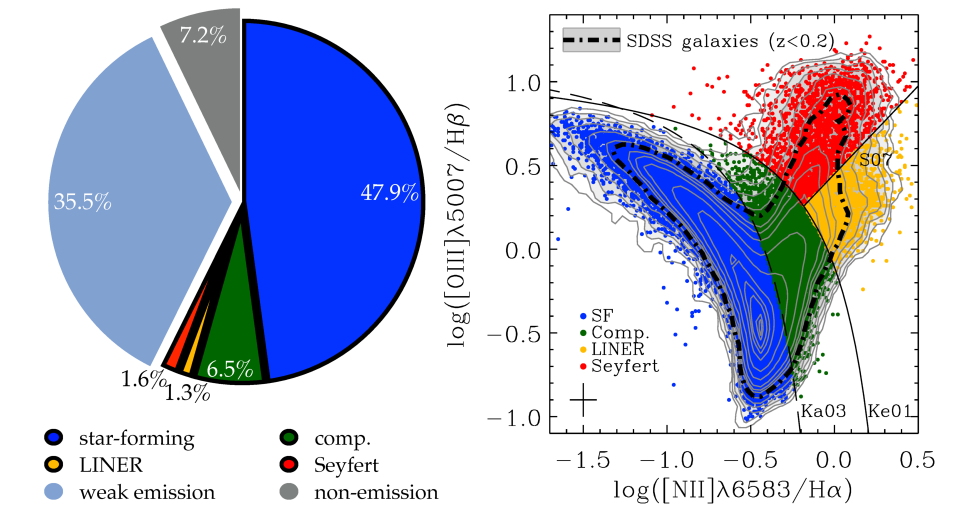}
\caption{Pie chart describing the parent samples (left panel) and
 the \NIIHa\ diagnostic diagram (right panel).
Left panel: Outline of the parent samples, divided based on emission-line characteristics. 
The loci of star-forming galaxies, composite galaxies, LINERs, and Seyferts, 
which comprise $57.3\%$ of the parent samples, are displayed in the right panel. 
Right panel: Emission-line classification of the SDSS data using a line diagnostic diagram 
\citep{Baldwin81, Kewley01, Kauffmann03, Kewley06, Schawinski07}.
The monochromatic filled contours and gray outlines are derived 
from the entire OSSY catalog at $z<0.2$ ($N\sim200,000$)
and the follow-up broad-line AGN spectral line measurements.  
These represent samples of SDSS emission-line galaxies with A/N $\ge3$ for \NII, \Ha, \OIII, and \Hb\ \citep{Oh11, Oh15}. 
The thick black dot-dashed line indicates the distribution of 95\% of the SDSS emission-line galaxies. 
Color-filled symbols denote the samples used in this study, with the same A/N cut applied at $0.01<z<0.06$. 
Blue, green, yellow, and red-filled dots represent star-forming galaxies, composite galaxies, LINERs, and Seyferts, respectively. 
Typical uncertainties are shown in the bottom left corner. 
\label{fig:pie_bpt}}
\end{figure*} 

Among the many groundbreaking techniques and methods developed 
in the fields of artificial intelligence and machine-learning, 
Convolutional Neural Networks (CNNs) have been widely applied to astronomical image analysis.  
However, traditional approaches to galaxy classification have ultimately relied on 
detailed spectroscopic analyses and emission-line diagnostic diagrams to determine star-forming activity. 

In parallel, Vision Transformers (ViT, \citealt{Vaswani17, Dosovitskiy21}) have made significant contributions 
to various tasks \citep{Gao21, Gheflati22, Lin22, Tanzi22, Cao24, Parker24}. 
ViT is an image classification architecture that enables neural networks to capture key features 
underlying long-range elements distributed across images, both locally and globally.  
When pre-trained on large datasets, ViT models have been shown to outperform ResNet-based baselines 
while requiring substantially fewer computational resources during pre-training \citep{Dosovitskiy21}
ViT has also demonstrated superior performance over state-of-the-art CNNs, 
particularly when dealing with large datasets, such as those containing hundreds of millions of images. 

Combined with several practical advantages of the JPEG image format,  
such as computational efficiency, reduced storage requirements, 
and compatibility with standard deep learning frameworks optimized for natural image processing,
these compelling features make ViT a highly suitable and 
effective technique for galaxy image classification and analysis.  
This is especially relevant given the rapidly increasing volume of data, 
such as that expected from the LSST \citep{Ivezic19}, 
which is projected to produce 20 billion galaxy images over its 10-year operational period in the near future.

In this paper, we present an innovative approach to classifying star-forming galaxies in the local universe using ViT.
The remainder of this paper is organized as follows. 
In Section \ref{sec:data}, we provide a brief introduction to the data used in this study. 
In Section \ref{sec:method}, we describe the ViT method applied to the data. 
We present the results and discuss our findings in Section \ref{sec:result}. 
We discuss the limitations and caveats of this work in Section \ref{sec:caveats}. 
Finally, we briefly summarize our results in Section \ref{sec:summary}.   
We assume a cosmology with $h = 0.70$, $\Omega_{M} = 0.30$, and $\Omega_{\Lambda}=0.70$ throughout this work.

\begin{figure*}
\centering
\includegraphics[width=0.844\linewidth, angle=0]{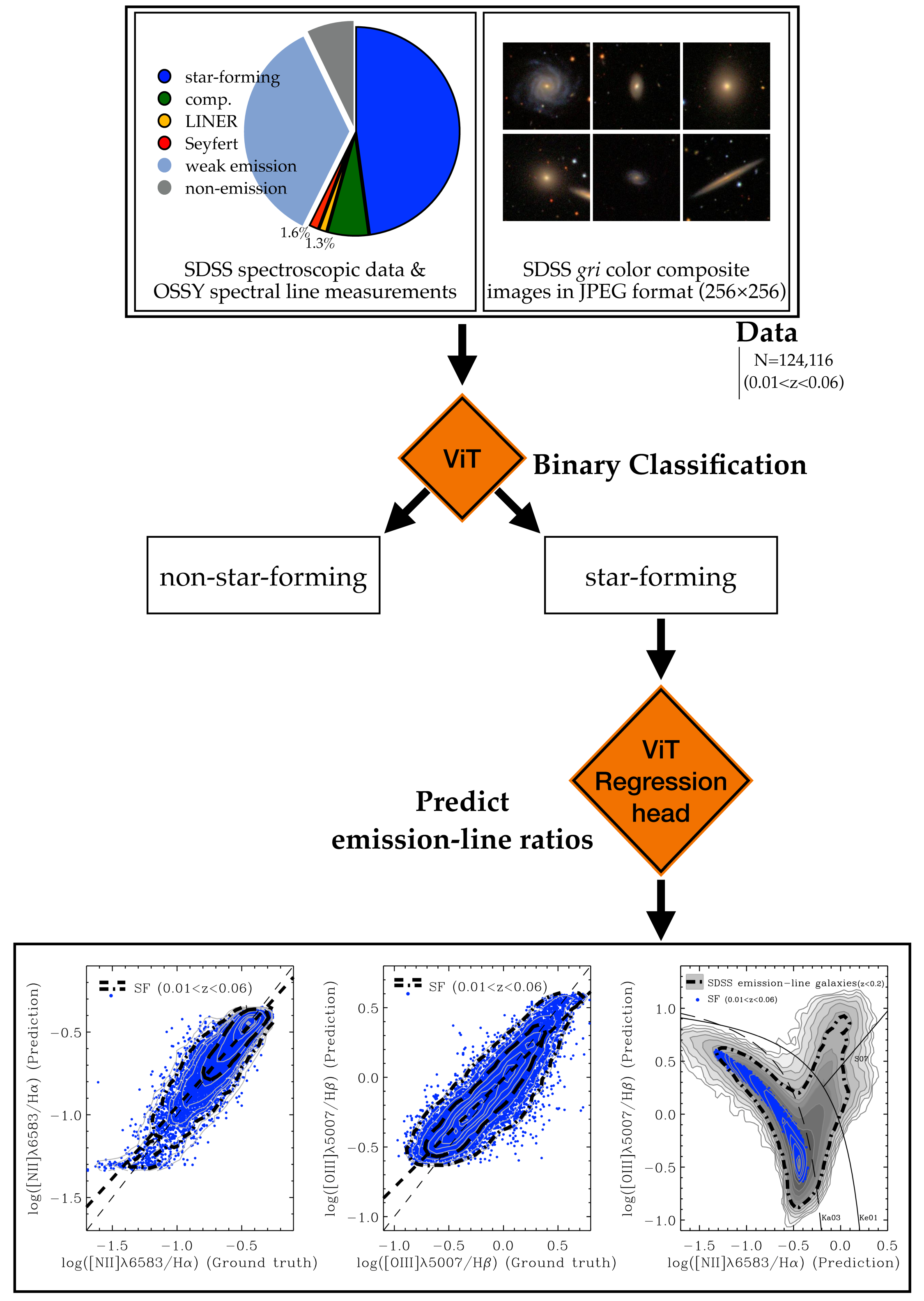}
\caption{Flowchart illustrating a general overview of this work. 
The Vision Transformer (ViT), shown in the middle with orange diamonds (Section~\ref{sec:method}), 
is fed SDSS \textit{gri} color composite images and used to predict emission line ratios. 
Section~\ref{sec:result} describes the results in detail.
\label{fig:flow_chart}}
\end{figure*} 

\section{Data} \label{sec:data}
We constructed our sample from the SDSS DR7 \citep{Abazajian09}, 
which represents the final and complete release of all spectroscopic and photometric data 
obtained during the SDSS Legacy Survey phase, 
and its follow-up spectroscopic line measurement database, 
the OSSY catalog \citep{Oh11}\footnote{\url{https://data.kasi.re.kr/vo/OSSY/}}. 
We began by selecting galaxies from the SDSS Main Galaxy Sample \citep{Strauss02}, 
a magnitude-limited sample defined by $r_{\rm petro} < 17.77$. 
To avoid sources affected by blending issues or large photometric errors
-- particularly in the case of extended galaxies --
we imposed a lower redshift limit and adopted an upper limit of $z=0.06$, 
ensuring that galaxy features could be examined in sufficient detail.

Specifically, our sample includes objects classified as `galaxies' by the SDSS pipeline, 
with spectroscopic redshifts in the range of $0.01<z<0.06$. 
Our final sample consists of \Nall\ galaxies, 
all of which were cross-matched with the Galaxy Zoo 1 data release\footnote{\url{https://data.galaxyzoo.org}} 
for morphology classification \citep{Lintott11}.

It is known that Galaxy Zoo morphological classifications can be biased 
due to varying user accuracy, inconsistent responses, and numbers of votes, which may contaminate the results. 
In this study, we did not apply a weighted-vote threshold for morphology classification, 
as the Galaxy Zoo morphology data are used solely to aid in interpreting potential causal relationships in morphology, 
and are not directly involved in the predictions of emission-line ratios.

In this study, we use the emission-line strengths measured by the OSSY catalog. 
The OSSY catalog is a spectroscopic database compiling galaxies from the SDSS Legacy Survey, 
comprising \NOSSY\ galaxies at $z<0.2$. 
It employs not only stellar population synthesis models \citep{Bruzual03} 
but also a set of empirical stellar templates based on the MILES stellar library \citep{Sanchez06}, 
enabling a more accurate continuum fit.

More importantly, the emission-line measurements provided by the OSSY database and 
the follow-up study by \citet{Oh15} are substantially more reliable in cases 
where broad-line components are present in the Balmer series. 
A significant number of type 1 AGNs, featuring weak broad-line regions in the local universe \citep{Oh15}, 
are known to be misclassified as narrow emission-line galaxies by the SDSS pipeline. 
For these weak AGNs, standard spectral line measurements often yield inaccurate emission-line strengths 
and ratios due to the presence of broad Balmer lines (see Figure 5 in \citealt{Oh11}). 
To address this, we use improved line measurements for these broad-line AGNs ($N=1,279$ at $0.01<z<0.06$) from \citet{Oh15}. 
In that work, an initial assumption of a broad Gaussian template with a width of $1,000$ \kms\ was imposed, 
enabling the detection of low-luminosity type 1 AGNs in the local universe 
and successfully decomposing the narrow lines from the broad components present in the \Ha\ and/or \Hb\ spectral complexes.

Our sample of SDSS galaxies is comprised as follows. 
Based on the Gaussian amplitude-to-noise ratios (A/N) from the spectroscopic line measurements, 
we first labeled emission-line galaxies as those with ${\rm A/N}\ge3$ for \NII, \Ha, \OIII, and \Hb, 
yielding \Nemission\ ($57.3\%$), which are placed into the \OIIIHb\ versus \NIIHa\ diagnostic diagram (blue, 
green, orange, and red wedges in the left panel of Figure~\ref{fig:pie_bpt}). 

We use the \OIIIHb\ versus \NIIHa\ diagnostic diagram because the four emission lines 
involved are the strongest in the optical spectrum of nearby galaxies and are, 
by definition, insensitive to dust extinction due to their small wavelength separation. 
This also facilitates securing a sufficient number of objects for training and testing. 

Weak-emission galaxies are defined as those with $1\le{\rm A/N}<3$ 
for any of the four lines ($N=\Nweak$, $35.5\%$, light-blue wedge), 
which do not meet the threshold to be placed on the BPT diagram. 
Lastly, non-emission galaxies are identified as those with ${\rm A/N}<1$ 
for all four emission lines simultaneously ($N=\Nnonemission$, $7.2\%$, gray wedge). 

Figure~\ref{fig:pie_bpt} illustrates our parent samples in terms of 
emission-line characteristics (left panel) and the BPT classifications (right panel). 
Our emission-line sample (${\rm A/N}\ge3$) is divided into four distinct classes based on the demarcation lines 
introduced by previous works \citep{Kewley01, Kauffmann03, Kewley06, Schawinski07}: 
star-forming galaxies ($N=\NSF$, $47.9\%$), 
composite galaxies ($N=\Ncomp$, $6.5\%$), 
LINERs ($N=\NLINER$, $1.3\%$), and Seyferts ($N=\NSY$, $1.6\%$). 

For our machine learning analysis, 
we group composite galaxies, Seyferts, LINERs, weak-emission galaxies, 
and non-emission galaxies together as a single `non-star-forming' class, 
creating a balanced binary classification task with star-forming galaxies 
representing $47.9\%$ and non-star-forming galaxies representing $52.1\%$ of the sample.

\begin{figure*}
\centering
\includegraphics[width=1.0\linewidth, angle=0]{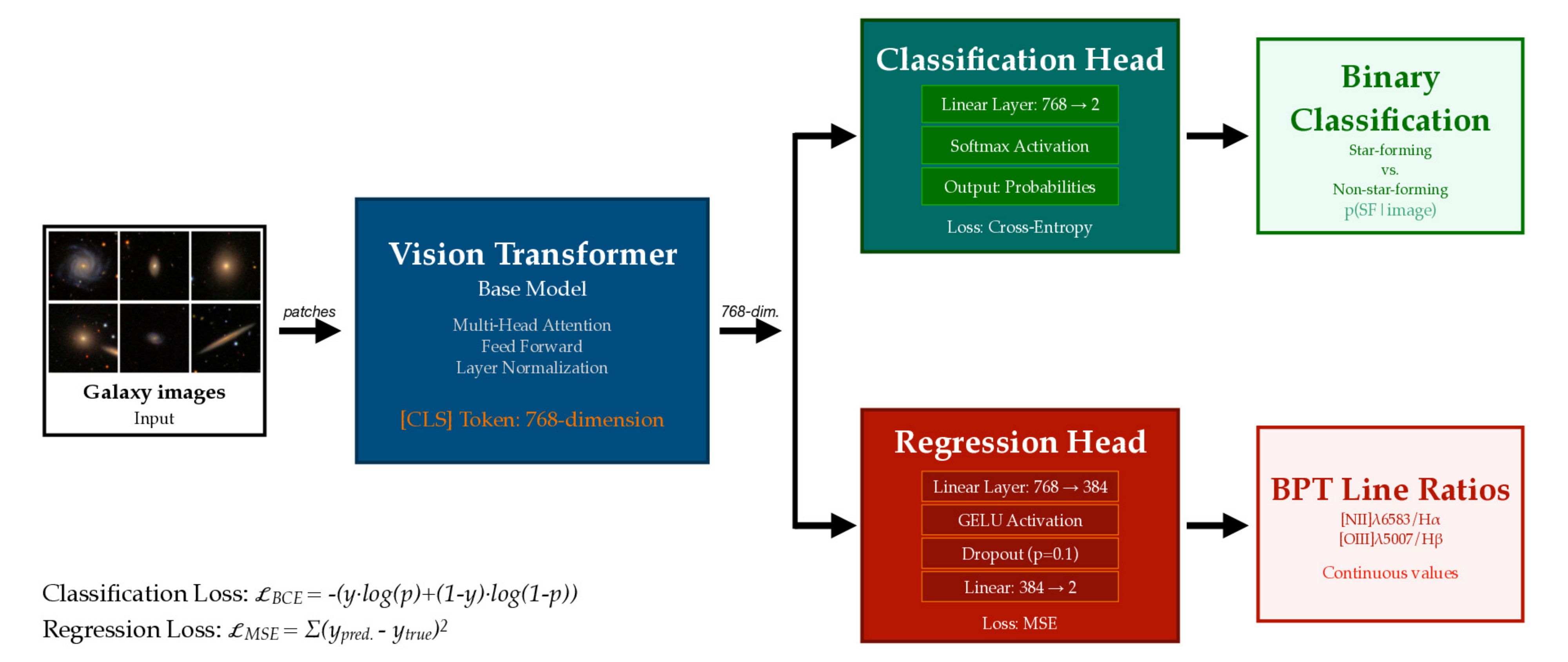}
\caption{Vision Transformer architecture for galaxy classification and regression. 
\label{fig:schematic_diagram_ViT}}
\end{figure*} 

\section{Method} \label{sec:method}
Our approach employs a Vision Transformer (ViT) architecture 
to analyze galaxy images for two distinct tasks: 
the classification of galaxies into star-forming and non-star-forming categories, 
and the prediction of emission line ratios. 
The following sections detail our implementation and modifications for these astronomical applications, 
illustrated in the flowchart shown in Figure~\ref{fig:flow_chart}.

\subsection{Vision Transformer Base Architecture} \label{sec:vit}
The foundation of our method is the Vision Transformer (ViT) model, 
a class of deep learning models that have achieved state of the art performance 
in a variety of image related tasks \citep{Dosovitskiy21}. 
The main idea behind Vision Transformers is to apply the transformer architecture, 
originally designed for natural language processing (NLP), 
to image data. This means processing images through a sequence of self-attention mechanisms 
rather than traditional convolutional operations. 
We utilize the pre-trained ViT-base model (patch size $16\times16$, image size $224\times224$) 
initially trained on ImageNet-21k and adapt it through transfer learning for our use case.

ViT-base strikes an appropriate balance between representational power and 
computational requirements for our dataset size and research objectives. 
Given that we are attempting to extract precise spectroscopic information from visual features, 
a challenging task that requires learning subtle morphological-spectroscopic correlations, 
ViT-base provides adequate complexity for this investigation.

The base model first segments each galaxy image into $16\times16$ pixel patches, 
projecting each patch into a 768-dimensional embedding space. 
These embeddings, combined with learned positional encodings, 
preserve both local features and spatial relationships within the galaxy images. 
A specialized classification token ([CLS]) aggregates information across all patches 
through the self-attention mechanism, providing a global representation of the galaxy's features.

The transformer backbone consists of 12 sequential blocks, 
each implementing multi-head self-attention followed by a feed-forward network. 
Layer normalization and residual connections are employed 
throughout to facilitate training and information flow. 
This architecture enables the model to capture both local features 
(such as star-forming regions or spiral arms) and global morphological characteristics simultaneously.

\subsection{Task-specific Implementations} \label{sec:implementation}
We develop two distinct model variants for our astronomical applications (Figure~\ref{fig:schematic_diagram_ViT}): 
a binary classifier for star-forming galaxy identification and a regression model for BPT line ratio predictions.

\subsubsection{Star-Formation Classification Variant}
For classifying galaxies into star-forming and non-star-forming categories, 
we implement a classification approach that operates on the 
768-dimensional [CLS] token representation from the vision transformer. 
The classification mechanism consists of a linear transformation that 
projects this representation to two output logits, 
corresponding to the two classes. 
The model is trained using cross-entropy loss, and during inference, 
a softmax function transforms these logits into class probabilities, 
where the probability for the star-forming class 
indicates the model's confidence in the galaxy being star-forming.

We use a cross entropy loss for the training process:
\begin{equation}
\mathcal{L}_{BCE}=-(ylog(p)+(1-y)log(1-p))
\end{equation}
where y is the true value of the galaxy being star-forming and p is the output probability by the model.

\subsubsection{BPT Line Ratio Regression Variant}
The regression variant, designed to predict BPT diagnostic line ratios (\NIIHa\ and \OIIIHb), 
maintains the same base model but is modified for continuous output values. 
The regression head processes the [CLS] token through an initial dimensionality reduction 
from 768 to 384 features, followed by GELU activation. 
A dropout layer ($p=0.1$) is employed for regularization. 
The final linear layer outputs two values, corresponding to the two line ratios. 
This architecture maintains sufficient complexity to capture the relationship 
between visual features and spectroscopic properties 
while preventing overfitting through dimensional reduction and dropout regularization.

We use the mean squared error as a loss for the training process:
\begin{equation}
MSE=\frac{1}{2}\sum\limits_{i=1}^2 (x_i - y_i)^2
\end{equation}
where $x_{0}$ is the predicted \NIIHa, $x_{1}$ is the predicted \OIIIHb, $y_{0}$ is the true \NIIHa, and $y_{1}$ is the true \OIIIHb.

\begin{figure}
\centering
\includegraphics[width=1\linewidth, angle=0]{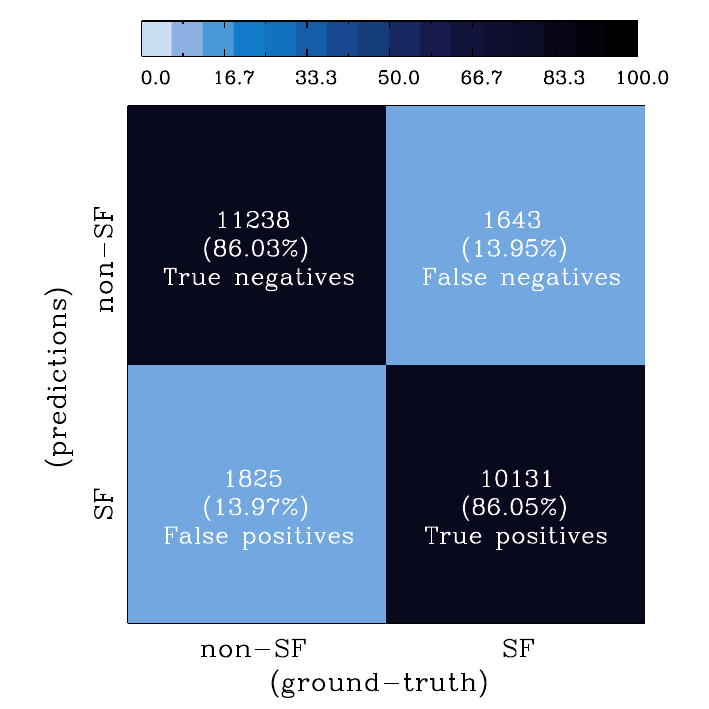}
\caption{Confusion matrix illustrating the results of the binary classification 
between star-forming galaxies and non-star-forming galaxies on the held-out $20\%$ test set. 
The number of sources in each category is displayed, with the corresponding fraction in parentheses and category labels. 
The four categories are color-coded according to the fraction, as indicated by the color bar. 
\label{fig:confusion_matrix}}
\end{figure} 

\begin{figure*}
\centering
\includegraphics[width=0.86\linewidth, angle=0]{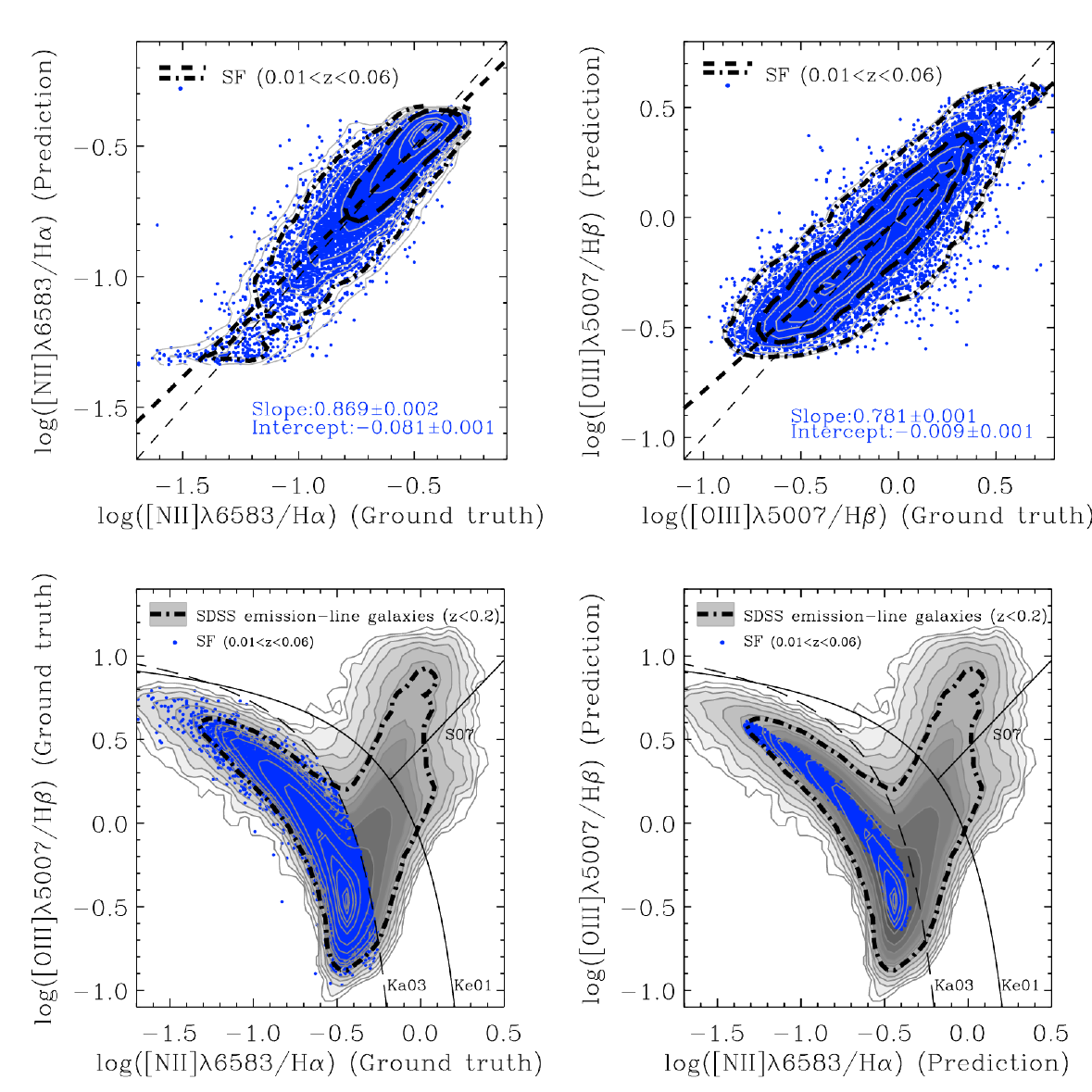}
\caption{Upper-panels: Comparison of emission-line ratios (predictions versus ground truth).
The test dataset, selected from the star-forming galaxies ($N_{\rm SF}=\NtrainingSF$), is used.
The upper-left and upper-right panels show the \NIIHa\ and \OIIIHb\ line ratios, respectively. 
Predicted emission-line ratios for star-forming galaxies (blue-filled dots) are compared with their observed values. 
Black dashed lines indicate the one-to-one fiducial relations. 
Thick black dashed contours and dot-dashed contours represent the 68\% and 95\% distributions, respectively. 
Linear regression fits are shown with thick black dashed lines, along with their corresponding slopes and intercepts. 
Lower-panels: Comparison of the \NIIHa\ diagnostic diagram (lower-left: ground truth; lower-right: prediction). 
The format is the same as in the right panel of Figure~\ref{fig:pie_bpt}.
\label{fig:bpt_pred_orig}}
\end{figure*} 

\subsubsection{Training Procedure}
Both model variants are initialized with weights from 
the pre-trained ViT-base model (google/vit-base-patch16-224), 
leveraging transfer learning to adapt to astronomical imagery. 
The models process galaxy images normalized to a standard $224\times224$ pixel format, 
maintaining consistency with the pre-trained architecture while providing sufficient resolution. 
We employ a standard 80-20 train-test split for all experiments, 
and we did not implement stratified samples as the natural distribution already ensures balanced representation in both training and test sets. 
In addition we evaluated our results on entirely independent samples of galaxies from different redshift ranges ($0.06<z<0.07$, $0.16<z<0.17$) 
as seen in Section~\ref{ssec:validation}.
Data augmentation techniques, including random rotations and flips, 
are employed to enhance model robustness to orientation variations.
For the regression task of line ratio prediction, 
we train for 20 epochs using a learning rate of $2\times10^{-4}$ with a cosine decay schedule 
and $10\%$ warmup period, building on established practices for transformer-based models \citep{Devlin19, Dosovitskiy21}. 
The classification model for identifying star-forming galaxies is trained for 
20 epochs with a learning rate of $5\times10^{-4}$. 
Both models employ gradient accumulation over 4 steps, 
effectively increasing the batch size while managing GPU memory constraints.
For regularization, we apply weight decay of 0.01 and gradient clipping 
with a maximum norm of 1.0 to the regression model. 
Both models utilize mixed-precision training (FP16) following \citet{Micikevicius18}, 
which optimizes memory usage and training speed while maintaining numerical stability. 
Model checkpoints are saved and evaluated every 100 steps, 
with the best performing model selected based on the 
relevant metric for each task - root mean square error (RMSE) for regression and accuracy for classification.

\section{Result and Discussion} \label{sec:result}
\subsection{Star-Formation Classification} \label{sec:sf_classification}
Figure~\ref{fig:confusion_matrix} presents the confusion matrix 
for the binary classification of star-formation galaxies on the held-out 20\% test set.
Both true positives and true negatives, 
which represent correctly classified star-forming and non-star-forming galaxies, 
achieve accuracies of approximately $86\%$. 
The remaining $\sim$$14\%$ of cases are classified as false positives or false negatives, 
indicating that the probabilities of misclassification as either star-forming or 
non-star-forming are nearly equal. 
The model also achieves a precision of 0.85, recall of 0.86, and F1 score of 0.85.
We discuss the possible origins of false positives and false negatives in 
Section~\ref{sec:false_pos_false_neg} and Figure~\ref{fig:binary_classification_failure}, 
as well as galaxy morphologies and apparent sizes in Section~\ref{sec:morphology_size} and Figure~\ref{fig:morphology_size}.

\subsection{Emission-Line Ratios} \label{sec:emission_line_ratio}
As illustrated in Figure~\ref{fig:flow_chart}, 
we used star-forming galaxies that passed the binary classification 
to predict their emission-line ratios in the BPT diagnostic diagram, 
shown in Figure~\ref{fig:bpt_pred_orig}. 
Both \NIIHa\ and \OIIIHb\ line ratios, shown in the top panels of Figure~\ref{fig:bpt_pred_orig}, 
exhibit a reasonably strong linear relationship, 
yielding coefficients of determination ($R^{2}$) of 0.837 and 0.830, respectively.

We obtained a Root Mean Squared Error (RMSE) of 0.089, 
indicating the typical magnitude of the prediction error, 
and a Mean Absolute Error (MAE) of 0.065, 
representing the average absolute deviation from the ground-truth 
for the \NIIHa\ line ratio. Slightly higher values were observed 
for the \OIIIHb\ line ratio, with an RMSE of 0.161 and an MAE of 0.124.
The error distributions, along with the RMSE and MAE, are presented in Section~\ref{sec:error_distribution} and Figure~\ref{fig:error_distribution}.
The offsets observed in both line ratios arise from the fact that 
the majority of the galaxies are located near the region 
where \NIIHa\ and \OIIIHb\ peak around $(-0.45, -0.45)$, 
as clearly shown in the bottom panels of Figure~\ref{fig:bpt_pred_orig}. 
Due to this disproportionate distribution in each emission-line ratio, 
the prediction tends to be skewed toward the peak values 
when the extract features are insufficient to correlate them.

\begin{figure*}
\centering
\includegraphics[width=1\linewidth, angle=0]{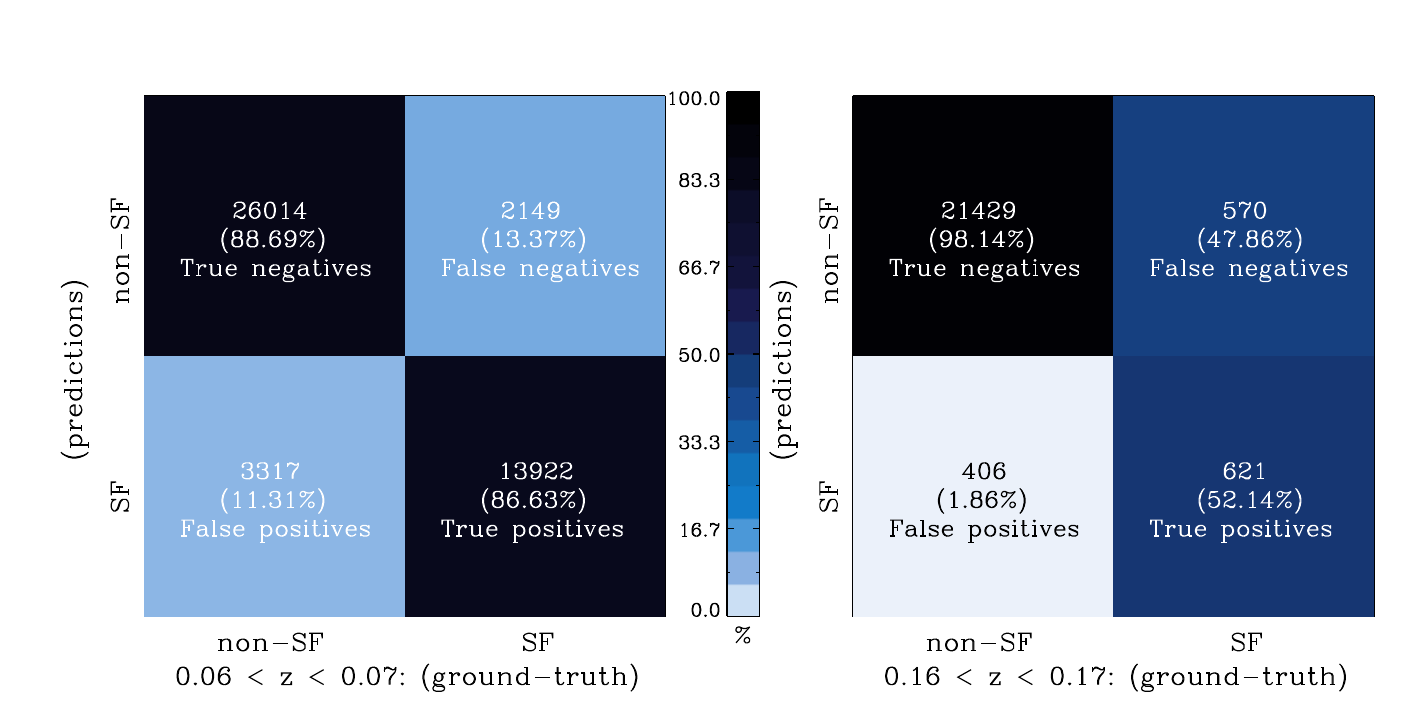}
\caption{Confusion matrix illustrating the results of the binary classification 
between star-forming galaxies and non-star-forming galaxies, 
applied to the evaluation dataset in the redshift range $0.06 < z < 0.07$ (left panel) and $0.16<z<0.17$ (right panel). 
The format is consistent with that of Figure~\ref{fig:confusion_matrix}.
\label{fig:confusion_matrix_06_z_07}}
\end{figure*} 

This disparity and the tendency exhibited in the ViT architecture 
also hold for the distribution in the 2-dimensional plane of the \OIIIHb\ versus \NIIHa\ diagram. 
The dispersion along the star-forming sequence, 
which stretches from a high \NIIHa\ and low \OIIIHb\ to a low \NIIHa\ and high \OIIIHb\ regime, 
becomes smaller in the prediction, resulting in a narrower ridge distribution in the diagnostic diagram. 
For a given \NIIHa\ ratio, the ViT architecture tends to predict the emission-line ratio 
most likely to fall within the main stream of the population. 
This is also true for the \OIIIHb\ ratio, 
which naturally leads to a smaller dispersion in the predicted diagnostic diagram 
(bottom-right panel of Figure~\ref{fig:bpt_pred_orig}).  

We have demonstrated how our ViT architecture predicts the emission-line ratios of star-forming galaxies. 
For galaxies classified as star-forming, 
the ViT architecture accurately describes the expected line ratios in the BPT diagnostic diagram.
However, this initial application of our ViT architecture to star forming galaxies does not provide a broader view of the heterogeneous nature of galaxies.
To address this, we discuss the prediction of emission-line ratios in the BPT diagram 
for all galaxy types (i.e., star-forming, composite, Seyfert, and LINER)
in Section~\ref{sec:predict_all_data}, along with Figure~\ref{fig:bpt_pred_orig_all}.

\begin{figure*}
\centering
\includegraphics[width=0.92\linewidth, angle=0]{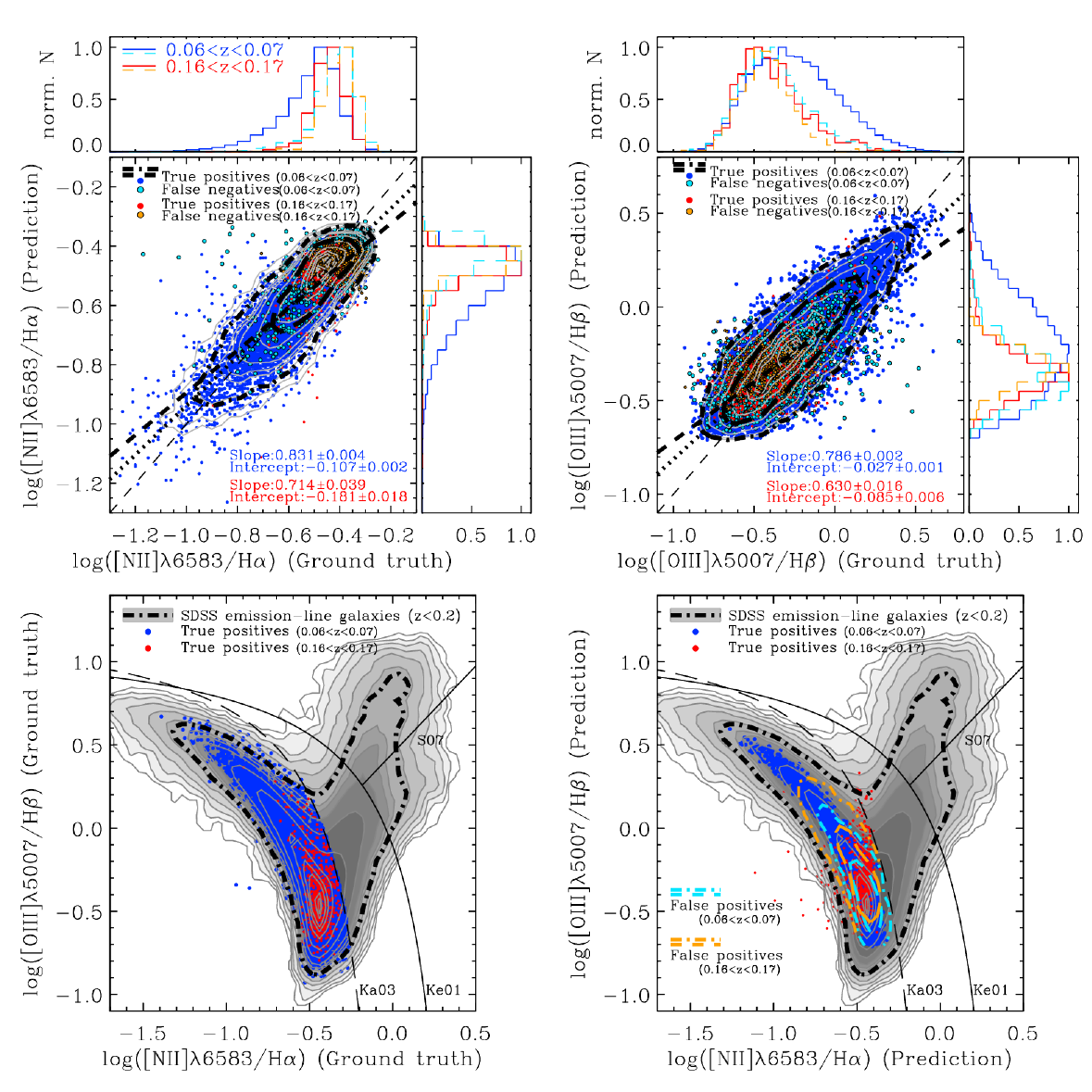}
\caption{Upper-panels: Comparison of emission-line ratios (predictions versus ground truth) for the test set.
The upper-left and upper-right panels show the \NIIHa\ and \OIIIHb\ line ratios, respectively. 
Predicted emission-line ratios for star-forming galaxies are indicated with blue-filled dots for $0.06<z<0.07$ ($N_{\rm SF,0.06<z<0.07}=\NtestSFTP$) 
and red-filled dots for $0.16<z<0.17$ ($N_{\rm SF,0.16<z<0.17}=\NtestSFhzTP$), and are compared with their observed values. 
Black dashed lines indicate the one-to-one fiducial relations. 
Black dashed contours represent the 68\% and 95\% distributions for the true positives at $0.06<z<0.07$. 
Linear regression fits are shown with dotted lines ($0.06<z<0.07$) and dashed lines ($0.16<z<0.17$), 
with their corresponding slope and intercept labels. 
False positives from both redshift bins are shown as light blue-filled dots and orange-filled dots.
Lower-panels: Comparison of the \NIIHa\ diagnostic diagram for both redshift bins (lower-left: ground truth; lower-right: prediction).
False positives from both redshift bins are shown as light blue contours (dashed for $68\%$, dot-dashed for $95\%$) and 
yellow contours. The format follows that of the right panel of Figure~\ref{fig:pie_bpt}.
\label{fig:bpt_06_z_07}}
\end{figure*} 

\subsection{Performance Validation on Unseen Redshift Ranges} \label{ssec:validation}
To evaluate our model's ability to generalize beyond its training distribution, 
we tested our ViT architecture on an extrapolation dataset 
containing galaxies with redshifts outside the training range. 
While our model was trained on galaxies within the redshift range of $0.01-0.06$, 
this evaluation dataset consists of galaxies with two redshift bins at $0.06 < z < 0.07$ $(N=\Ntestalllow)$ 
and $0.16 < z < 0.17$ $(N=\Ntestallhi)$, where the SDSS and the OSSY database provide enough number of galaxy photometric images and spectroscopic data. 
This experimental design allows us to assess whether the learned features remain valid when applied to a dataset that has never been exposed to the model. 

As presented in Section~\ref{sec:sf_classification}, 
the confusion matrix in Figure~\ref{fig:confusion_matrix_06_z_07} demonstrates 
the model's performance on this unseen data distribution. 
For the evaluation dataset in the low-z (left panel in Figure~\ref{fig:confusion_matrix_06_z_07}), first of all, 
the fractions of true positives $(\sim$$86.6\%)$ and 
true negatives $(\sim$$88.7\%)$ are slightly higher than those observed in 
our previous validation experiments, 
suggesting robust generalization capabilities.
On this extrapolated dataset, we report a precision of 0.81, recall of 0.87, and F1 score of 0.84.

The results at higher redshifts ($0.16 < z < 0.17$), shown in the right panel of Figure~\ref{fig:confusion_matrix_06_z_07}, 
reveal significantly degraded classification performance compared to our validation results at lower redshifts 
(c.f., a precision of 0.60, recall of 0.52, and F1 score of 0.56).
The confusion matrix indicates a substantially higher false negative rate ($47.86\%$ versus $13.37\%$) 
and a reduced true positive rate ($52.14\%$ versus $86.63\%$), 
suggesting that the model's learned features do not generalize effectively to this redshift regime.

This performance degradation can be attributed primarily to observational and methodological factors. 
Most notably, galaxies at $0.16 < z < 0.17$ appear substantially smaller in angular size 
than those in the low-redshift training sample, fundamentally altering the spatial scale at which emission-line features are resolved. 
Our Vision Transformer architecture, trained predominantly on nearby galaxies with larger apparent sizes and well-resolved morphological structures, 
may struggle to extract meaningful features from these more compact, lower surface brightness sources. 
The model's tendency to classify higher-redshift galaxies as non-star-forming 
(as evidenced by the high false negative rate) is consistent with these observational limitations. 

When star-forming regions are poorly resolved or when emission-line features are diluted 
due to reduced angular size and lower surface brightness, 
the model likely defaults to the more conservative non-star-forming classification. 
This bias toward non-star-forming predictions in the presence of ambiguous or 
unresolved features likely reflects the model's training on clearer, 
more distinct examples at lower redshifts. 
At these higher redshifts, the combination of reduced angular resolution and lower signal-to-noise ratios contributes 
to the observed classification challenges, 
as the model encounters spatial scales and data quality conditions that differ substantially from those present in the training set.

The top panels of Figure~\ref{fig:bpt_06_z_07} present the \NIIHa\ and \OIIIHb\ line ratios of star-forming galaxies, 
as shown in Figure~\ref{fig:bpt_pred_orig}, 
while further differentiating true positives (blue-filled dots at $0.06 < z < 0.07$, $\sim$$87\%$, and red-filled dots at $0.16 < z < 0.17$, $\sim$$52\%$) 
and false negatives (light blue-filled dots for $0.06 < z < 0.07$, $\sim$$13\%$, and orange-filled dots for $0.16 < z < 0.17$, $\sim$$48\%$) 
within the star-forming galaxy population. 
The false negatives are primarily located in the regions where the \NIIHa\ and \OIIIHb\ ratios are most frequently distributed.  

The combined emission-line ratios in the BPT diagram, shown in the lower panels of Figure~\ref{fig:bpt_06_z_07}, 
well reproduce the observed values, although with less dispersion along the star-forming sequence. 
The $68\%$ and $95\%$ distributions of false positives (light blue contours for $0.06 < z < 0.07$, $\sim$$11\%$, 
and orange contours for $0.16 < z < 0.17$, $\sim$$2\%$), which were classified as star-forming 
but are actually non-star-forming, are overlaid on top of the true positives (blue-filled dots and red-filled dots), 
closely mimicking the densest region of the sequence. 

The majority of star-forming and non-star-forming galaxies are correctly classified 
in the validation data drawn from the low redshift bin 
(with $\sim$$87\%$ for true positives and $\sim$$89\%$ for true negatives).  
However, the classification performance deteriorates at higher redshift, as discussed earlier. 

The false negatives (light blue-filled dots and orange-filled dots) appear at $\sim$$13\%$ and $\sim$$48\%$ in the two redshift bins, respectively. 
The false positives (light blue contours and orange contours) appear at $\sim$$11\%$ and $\sim$$2\%$, respectively.
The locations of these misclassifications in the BPT diagram indicate that 
our ViT architecture generally predicts emission-line ratios of star-forming galaxies well.  
The model performs especially well in the region of low-\NIIHa\ and high-\OIIIHb\ values, where low-mass, metal-poor galaxies tend to lie, 
as described in Figure 1 of \cite{Kumari21} and references therein. 
However, the performance declines in the higher redshift regime.

\section{Limitations and Caveats} \label{sec:caveats}
Our current approach relies on transfer learning from a ViT model pre-trained on 
large-scale natural image datasets, which have proven effective for astronomical applications. 
The morphological features distinguishing star-forming galaxies 
(e.g., spiral arms, clumpy structures, overall shapes) share structural similarities with features learned 
from natural images, making this approach well-suited to our task of predicting spectroscopic properties from apparent galaxy features.

However, this strategy presents inherent limitations in astronomical applications as well. 
The features learned from natural images --  such as edges, textures, and object boundaries -- 
may not optimally represent the unique characteristics of galaxy morphology. 
Natural images typically contain sharp edges, distinct objects, and structured backgrounds, 
whereas galaxy images exhibit smooth intensity gradients, 
complex hierarchical structures, and noise properties specific to astronomical observations.

Another fundamental limitation of our preprocessing pipeline is 
the conversion of astronomical FITS images to JPEG format. 
This transformation, while necessary for compatibility with standard computer vision architectures, 
discards the precise flux measurements present in the original FITS files. 
The limited dynamic range of the JPEG format results in the loss or dimming of potentially important information, 
such as bright emissions from the very center of a galaxy due to AGN activity.

It is evident that our ViT architecture performs inadequately 
when the images lack sufficient information. 
This fundamental (and most likely causal) limitation significantly contributes to the challenges 
in predicting various spectroscopic classes of galaxies dominated by nuclear emission,  
as discussed in Section~\ref{sec:predict_all_data} and shown in Figure~\ref{fig:bpt_pred_orig_all}. 
When the input galaxy image lacks distinct features, as in the case of Seyferts and LINERs, 
the predicted emission-line ratios and their corresponding locations in the BPT diagram become inaccurate, 
biased toward the most frequently observed values in the ground truth. 
This issue arises because nuclear emissions occupy only a tiny fraction of the given JPEG image space 
-- often less than a single pixel -- naturally explaining both the failure in predicting emission-line ratios of 
Seyferts (red-filled dots in Figure~\ref{fig:bpt_pred_orig_all}) and LINERs (orange-filled dots) 
and the relative success in composite galaxies (green-filled dots).  
It is noteworthy that the Seyfert galaxies in this study, drawn from the SDSS, 
represent nearby weak AGNs with bolometric luminosities in the range 
$10^{43}<L_{\rm bol}<10^{46}$ \ergs\ \citep[see Figure 2, Figure 5, and Figure 9 in][]{Oh15}. 
Unlike powerful high-redshift quasars, these Seyferts do not exhibit outshining emission across their host galaxies.  

It should be addressed that the marginally better performance observed 
in the extrapolated redshift range of $0.06 < z < 0.07$ merits careful interpretation. 
While this result may initially suggest improved generalization, 
it is more plausibly attributed to statistical fluctuations and the proximity of this range 
to the upper bound of the training distribution ($z = 0.06$), 
rather than a true enhancement in model capability. 
This interpretation is substantiated by the model’s markedly poorer performance at higher redshifts ($0.16 < z < 0.17$), 
where the true positive rate declines significantly (from $86.63$\% to $52.14$\%), 
as shown in the right panel of Figure~\ref{fig:confusion_matrix_06_z_07}. 
This clear degradation highlights the model’s limitations in extrapolation and 
suggests that the performance at $0.06 < z < 0.07$ remains within the expected variation of validation-like data. 
Therefore, the slight improvement in this narrow redshift band should not be over-interpreted 
but instead viewed as consistent with statistical noise near the training boundary.

Finally, we note that our approach relies on standard RGB JPEG images and Vision Transformer architectures. 
While this enables broad accessibility and compatibility with existing deep learning tools, 
it also introduces limitations related to lossy compression and a lack of domain-specific optimization. 
Future work could explore more specialized alternatives, such as directly incorporating FITS-format data, 
adopting astronomy-specific pre-trained models, or implementing custom token-embedding schemes 
tailored to the statistical and structural characteristics of astronomical images. 
These directions would represent substantial methodological extensions and are beyond the scope of the present study.

\section{Summary} \label{sec:summary}
Using a Vision Transformer (ViT) base-model, 
we demonstrated that the spectroscopic features of star-forming galaxies in the 
\OIIIHb\ versus \NIIHa\ BPT diagnostic diagram can be predicted purely based on 
optical color composite JPEG images, 
establishing a direct link between apparent visual features and their underlying spectroscopic characteristics.
Applying this approach to a sample of approximately 124,000 nearby galaxies 
from SDSS in the redshift range $0.01 < z < 0.06$, 
the model achieved strong classification performance, with a precision of 0.85, recall of 0.86, and an F1 score of 0.85. 
In terms of emission-line ratio regression, 
the predicted \NIIHa\ and \OIIIHb\ values showed close agreement with spectroscopic ground truth, 
yielding root mean square errors (RMSE) of 0.089 and 0.161, and mean absolute errors (MAE) of 0.065 and 0.124, respectively.

The predicted distribution of star-forming galaxies in the BPT diagram 
closely follows the observed star-forming sequence, exhibiting reduced dispersion along the ridge. 
This pattern reflects the model’s tendency to favor predictions that align with the high-density regions of the sequence, 
likely corresponding to typical morphologies seen in low-redshift star-forming systems. 
Notably, the apparent angular size of galaxies in the JPEG images was not found to be a critical factor in classification accuracy. 
However, misclassifications, comprise approximately $14\%$ of the sample, were more frequently associated with galaxies 
having larger-than-median angular diameters. 
These galaxies typically occupied an intermediate size regime 
between the dominant populations of true positives and true negatives. 
Moreover, false negatives tended to be more massive than true positives, 
deviating from the expected stellar mass–star formation rate relation.

When evaluated on an entirely new and previously unseen dataset, 
the model demonstrated slightly improved generalization performance, 
achieving approximately $87\%$ true positives and $89\%$ true negatives. 
However, a clear decline in performance is observed 
when the validation dataset includes galaxies selected from higher redshifts.
Additionally, the model proves ineffective at 
predicting the emission-line ratios of Seyferts and LINERs using JPEG images alone, 
as the limited morphological information contained in these images is insufficient to capture the ionization conditions and 
distinguishing characteristics of these active galactic nuclei.

We hope that this study will be applicable to both current and future large-scale surveys, 
including Euclid \citep{Euclid}, the Dark Energy Survey \citep{Abbott18}, and the Legacy Survey of Space and Time \citep{Ivezic19}. 
We provide the code and the dataset used in this work to the community 
through the repository\footnote{\url{https://data.kasi.re.kr/vo/sf-from-jpeg/}}, 
with the aim of fostering further fruitful scientific applications. 

\begin{acknowledgments}
We thank Min-Su Shin and Kevin Schawinski for taking the time to give valuable advice. 

K.O. acknowledges support from the Korea Astronomy and
Space Science Institute under the R\&D program (Project No. 2025-1-831-02), 
supervised by the Korea AeroSpace Administration,  
and the National Research Foundation of Korea (NRF) grant funded by the Korea government (MSIT) (RS-2025-00553982). 

This research has made use of NASA’s ADS Service. 
This research has made use of the NASA/ IPAC Infrared Science Archive, 
which is operated by the Jet Propulsion Laboratory, California Institute of Technology, 
under contract with the National Aeronautics and Space Administration.

Funding for the Sloan Digital Sky Survey (SDSS) and SDSS-II has been provided by the Alfred P. Sloan Foundation, 
the Participating Institutions, the National Science Foundation, the U.S. Department of Energy, 
the National Aeronautics and Space Administration, the Japanese Monbukagakusho, 
and the Max Planck Society, and the Higher Education Funding Council for England. 
The SDSS Web site is http://www.sdss.org/.

The SDSS is managed by the Astrophysical Research Consortium (ARC) for the Participating Institutions. 
The Participating Institutions are the American Museum of Natural History, Astrophysical Institute Potsdam, 
University of Basel, University of Cambridge, Case Western Reserve University, The University of Chicago, 
Drexel University, Fermilab, the Institute for Advanced Study, the Japan Participation Group, 
The Johns Hopkins University, the Joint Institute for Nuclear Astrophysics, 
the Kavli Institute for Particle Astrophysics and Cosmology, the Korean Scientist Group, 
the Chinese Academy of Sciences (LAMOST), Los Alamos National Laboratory, 
the Max-Planck-Institute for Astronomy (MPIA), the Max-Planck-Institute for Astrophysics (MPA), 
New Mexico State University, Ohio State University, University of Pittsburgh, University of Portsmouth, 
Princeton University, the United States Naval Observatory, and the University of Washington.
\end{acknowledgments}

%





\appendix

\section{False positives and false negatives} \label{sec:false_pos_false_neg}

\begin{figure}
\centering
\includegraphics[width=1\linewidth, angle=0]{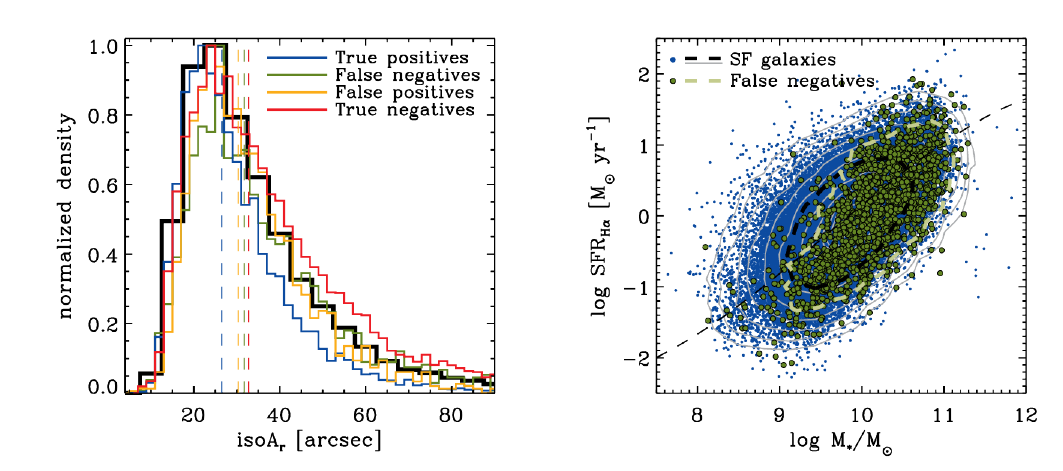}
\caption{Distribution of the angular diameter of galaxies, 
inferred from the SDSS isophotal major axis in the $r$ band (${\rm isoA_{r}}$), in units of arcseconds (left panel), 
and the star formation rate (SFR, in \Msunyr) as a function of stellar mass (right panel). 
Left panel: The four categories presented in Figure~\ref{fig:confusion_matrix} are illustrated with colored histograms, 
while the black histogram represents the combined populations. 
The median of each category is indicated by a vertical dashed line. 
Right panel: The SFR versus stellar mass for false negatives ---star-forming galaxies misclassified as non-star-forming---is shown as dark green-filled dots, 
overlaid on all star-forming galaxies, represented by blue-filled dots. 
The light green thick dashed line and black thick dashed line indicate the $68\%$ distribution of false negatives 
and all star-forming galaxies, respectively.
The black thin dashed line represents the relation between the SFR and stellar mass for star-forming galaxies drawn from the SDSS \citep{DuartePuertas17}.
\label{fig:binary_classification_failure}}
\end{figure} 

It is reasonable to infer that the apparent size of a galaxy, as presented in the SDSS \textit{gri} color composite image, 
can be a critical factor in enabling effective extraction of characteristic features and linking them to spectroscopic properties. 
In general, larger galaxies tend to yield better classification performance, 
likely due to the richer spatial and morphological detail preserved in their images. 

However, as shown in the left panel of Figure~\ref{fig:binary_classification_failure}, 
our ViT architecture demonstrates robust performance even for galaxies with angular diameters below the median.
Galaxies correctly classified with $\sim$$86\%$ accuracy show either lower (true positives) or 
higher (true negatives) abundances compared to the total population, under the same normalization, 
in the regime of large apparent sizes. 
In contrast, the remaining $\sim$$14\%$ of misclassified cases, consisting both false positives and false negatives, 
are concentrated in an intermediate range of angular diameters, 
where galaxy sizes deviate from the typical profiles of true positives and true negatives. 

Importantly, the size-dependent behavior is observed within the redshift range used for training ($0.01<z<0.06$) 
and persists in the adjacent low-redshift validation range ($0.06<z<0.07$). 
However, when the model is applied to significantly higher-redshift galaxies ($0.16<z<0.17$, see Section~\ref{ssec:validation}), 
the overall classification performance declines, likely due to the loss of resolved features in the images.
The degradation is attributed to a reduction in the amount of morphological and structural information captured in the JPEG images of distant galaxies, 
where smaller angular sizes and surface brightness dimming collectively limit the visual features available to the model. 
Unlike the performance variance related to size within a fixed redshift range, 
this redshift-dependent effect reflects a more fundamental limitation in the imaging data quality at higher redshift.

False negatives, which are star-forming galaxies misclassified as non-star-forming, 
are further analyzed in the right panel of Figure~\ref{fig:binary_classification_failure}, 
based on their star formation rate (SFR) and stellar mass. 
The SFR is estimated using extinction-corrected \Ha\ emission line strengths 
and is subsequently corrected for aperture effects following \citet{DuartePuertas17}. 
The stellar mass is derived using the formulae introduced by \citet{Bell03}, 
which incorporate the $g-r$ color and $r$-band absolute magnitudes. 
The false negatives, comprising approximately $14\%$ of the genuine star-forming galaxies, 
are generally located in the high-mass regime, 
deviating from the majority of the overall star-forming population. 
The classification architecture for identifying star-forming galaxies tends to select such massive objects as false negatives.

\section{Galaxy morphology and apparent size} \label{sec:morphology_size}
\begin{figure}
\centering
\includegraphics[width=1\linewidth, angle=0]{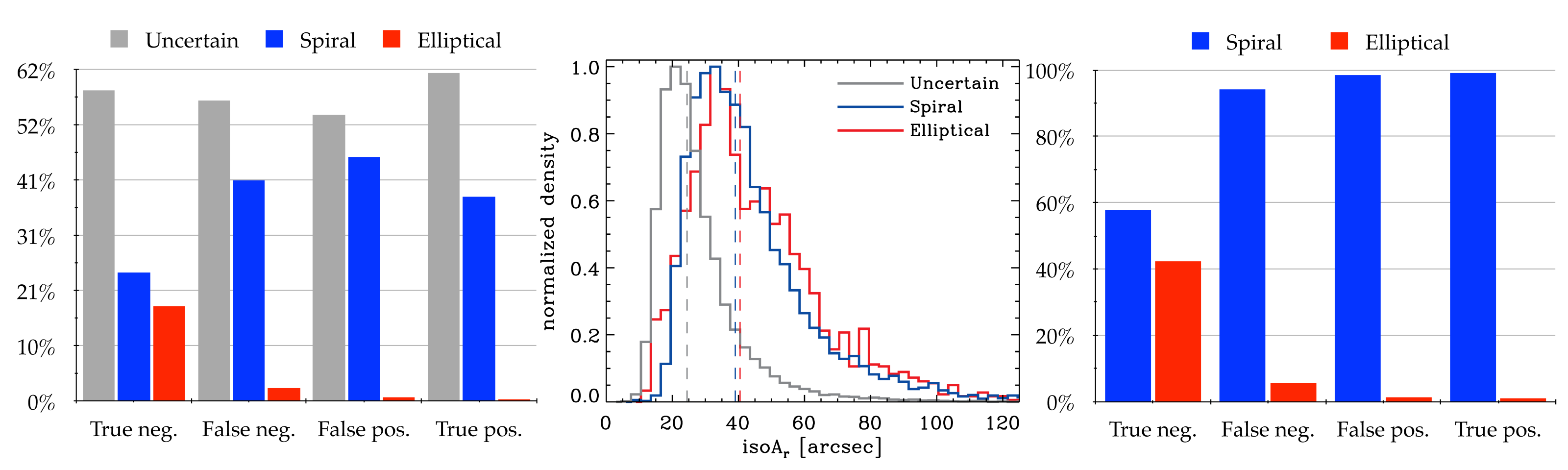}
\caption{Left panel: Morphology classifications from the Galaxy Zoo 1 data release \citep{Lintott11}, 
grouped according to each category in the confusion matrix presented in Figure~\ref{fig:confusion_matrix}. 
Middle panel: Distribution of apparent galaxy sizes, 
inferred from the SDSS isophotal major axis in the $r$ band (${\rm isoA_{r}}$), measured in arcseconds, 
for the three morphology classes shown in the left panel.  
Uncertain, spiral, and elliptical galaxies are represented in gray, blue, and red, respectively. 
Vertical dashed lines indicate the median values. 
Right panel: Same as the left panel, but excluding the uncertain category. 
\label{fig:morphology_size}}
\end{figure}

In Figure~\ref{fig:morphology_size}, we further analyze the four categories shown in Figure~\ref{fig:confusion_matrix} 
based on galaxy morphologies (left and right panels) and apparent sizes (middle panel). 
The Galaxy Zoo 1 data release \citep{Lintott11} provides a simple morphology classification scheme, 
categorizing galaxies as uncertain, spiral, or elliptical. 
The fraction of each morphology class within each category is presented in the left panel of Figure~\ref{fig:morphology_size}. 
Uncertain objects dominate all groups, comprising more than $50\%$, 
and they mostly have small apparent sizes (middle panel), 
making it difficult to accurately determine their true morphologies. 
As expected, the relative ratio of spiral to elliptical galaxies in each category increases 
from true negatives to true positives, as shown in the right panel of Figure~\ref{fig:morphology_size}, 
which excludes the uncertain objects presented in the left panel.

\section{Predicting emission-line ratios using all types of data} \label{sec:predict_all_data}
\begin{figure}
\centering
\includegraphics[width=0.86\linewidth, angle=0]{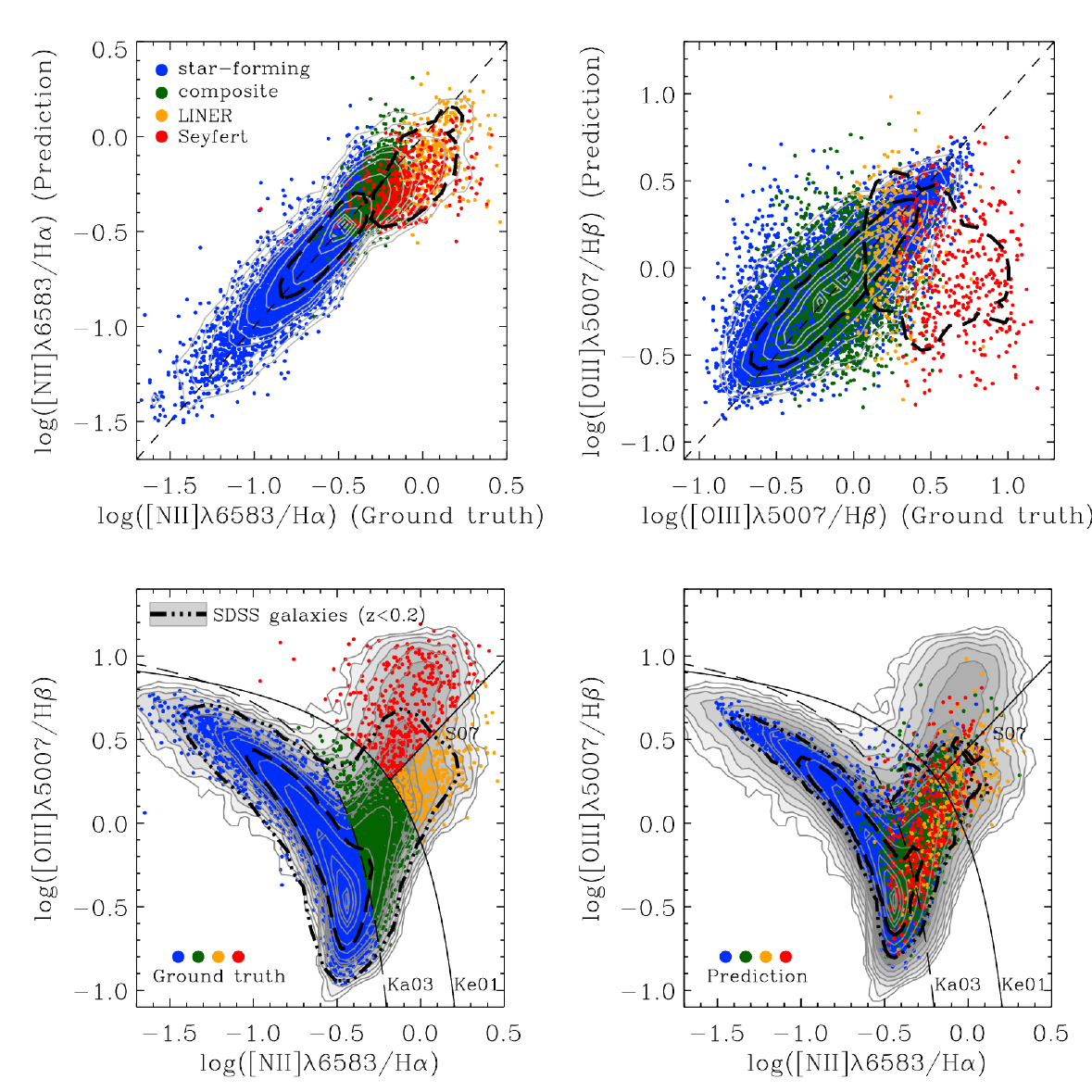}
\caption{Upper-panels: Comparison of emission-line ratios (predictions versus ground truth).
The test dataset, selected from the emission-line galaxies ($N_{\rm emi.}=\Ntestemi$), is used.
The upper-left and upper-right panels represent the \NIIHa\ and \OIIIHb\ line ratios, respectively. 
Predictions of emission-line ratios are compared with their observed values. 
The color-codes indicate star-forming galaxies (blue-filled dots), composite galaxies (green-filled dots), 
LINERs (orange-filled dots), and Seyferts (red-filled dots).  
Black dashed lines represent the one-to-one fiducial lines. 
Thick black dashed contours in each panel represent the 68\% distribution of star-forming galaxies (left contour) and 
the combined distribution of LINERs and Seyferts (right contour). 
Lower-panels: Comparison of the \NIIHa\ diagnostic diagram. 
The ground truth values (lower-left panel) and the predictions (lower-right panel) are shown with the corresponding color codes 
overlaid on the SDSS emission-line galaxies presented in the right panel of Figure~\ref{fig:pie_bpt}. 
Thick black dashed contours and dot-dashed contours in each panel represent the 68\% and 95\% distributions of all emission-line galaxies 
drawn from the test dataset. 
\label{fig:bpt_pred_orig_all}}
\end{figure}

Figure~\ref{fig:bpt_pred_orig_all} illustrates the \NIIHa\ and \OIIIHb\ emission-line ratios (upper panels) 
and their combined results in the BPT diagram (lower panels), 
as shown in Figure~\ref{fig:bpt_pred_orig} and Figure~\ref{fig:bpt_06_z_07}. 
The figure includes not only star-forming galaxies (blue-filled dots) 
but also other types of galaxies, such as composite (green-filled dots), 
LINERs (orange-filled dots), and Seyferts (red-filled dots). 

Since star-forming features are well extracted from the galaxy JPEG images, 
the predictions work for both star-forming and composite galaxies, 
which exhibit embedded star-forming features in the images. 
However, feature extraction for LINERs and Seyferts --  
where nuclear emissions play a key role in their classification in the BPT diagram -- 
is less effective due to the limited pixel coverage of the nucleus relative to the entire galaxy in the image. 
This lack of information biases the predictions toward the average values of \NIIHa\ and \OIIIHb, 
leading to spurious and unphysical results in the BPT diagram.

\section{Error Distribution} \label{sec:error_distribution}
\begin{figure}
\centering
\includegraphics[width=0.86\linewidth, angle=0]{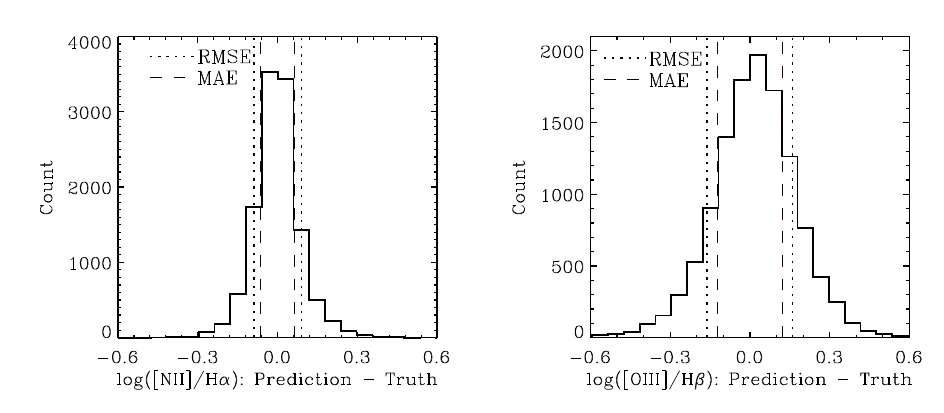}
\caption{Error distributions for the \NIIHa\ (left panel) and \OIIIHb\ (right panel) emission-line ratios. 
The dotted and dashed lines indicate the RMSE and MAE, respectively. 
\label{fig:error_distribution}}
\end{figure}

Figure~\ref{fig:error_distribution} presents the distribution of prediction errors, 
computed as the difference between the ViT model predictions and ground-truth values. 
The histograms in both panels peak near zero, indicating that the model is generally unbiased. 
The error distributions are fairly symmetric. 
The RMSE (vertical dotted line) and MAE (vertical dashed line) for \NIIHa\ is 0.089 and 0.065, respectively, 
while values of 0.161 and 0.124 are obtained for \OIIIHb, which are slightly larger than those for \NIIHa. 
Our ViT model predicts \NIIHa\ with higher accuracy than \OIIIHb, suggesting that the latter is more challenging to model, 
likely due to its greater physical complexity.

\bibliography{references}{}
\bibliographystyle{aasjournal}



\end{document}